\documentclass{evn2004}
\usepackage{graphicx}
\usepackage{txfonts}
\begin{document}
\title{First VLBI mapping of a rare SiO isotopic substitution}
   
   \author{R. Soria-Ruiz\inst{1} \and F. Colomer\inst{2} \and J.
     Alcolea\inst{1} \and V. Bujarrabal\inst{2} \and J.-F.
     Desmurs\inst{1} \and K.B. Marvel\inst{3}
          }

   \institute{ Observatorio Astron\'omico Nacional, c/Alfonso XII 3,
     28014 Madrid, Spain \and Observatorio Astron\'omico Nacional,
     Apartado 112, 28803 Alcal\'a de Henares, Spain \and American 
     Astronomical Society, 2000 Florida Avenue NW Suite, Washington,
     DC-20009-1231, USA}

   \abstract{We report the first VLBI map of the 7\,mm $v$=0 $J$=1--0
   maser line of the $^{29}$SiO isotopic substitution in the long
   period variable star IRC\,+10011. We have found that this maser
   emission is composed of multiple features distributed in an
   incomplete ring, suggesting that this maser is also amplified
   tangentially as already proposed in other SiO circumstellar
   masers. We present also the results for some $^{28}$SiO 7\,mm and
   3\,mm maser lines. We confirm in this epoch that the 86 GHz maser
   $v$=1 $J$=2--1 in IRC\,+10011 is formed in an outer region of the
   circumstellar envelope compared to the other SiO masers
   studied. The $^{29}$SiO masing region appears to be located in a
   layer in between the $^{28}$SiO $v$=1 $J$=1--0 and $^{28}$SiO $v$=1
   $J$=2--1 lines. Finally, we discuss the possible implications of
   the observational results in the SiO maser pumping theory.
  
 }

   \authorrunning{R. Soria-Ruiz et al.}
   \titlerunning{mm-VLBI observations of $^{29}$SiO in IRC+10011}

   \maketitle
%

\section{Introduction}

SiO maser emission is detected in the innermost shells of the
circumstellar envelopes around Long Period Variable (LPV) stars. 
Three different isotopic subtitutions of the SiO molecule
are known to present maser emission in these objects: 
$^{28}$SiO, $^{29}$SiO and $^{30}$SiO. 
The study of this type of emission provides information about the 
physical processes that occur near the central source.

Very long baseline interferometric (VLBI) observations have also
contributed to a better understading of the SiO maser phenomenon itself,
although only the strong 7\,mm and recently the 3\,mm rotational
transitions of $^{28}$SiO, which is the most abundant isotopic 
substitution, have been mapped in these LPVs (see for example 
Doeleman et al. 1998 and Desmurs et al. 2000). In fact, prior to
 this work, no VLBI detection of SiO maser lines other than 
those of the main isotope $^{28}$SiO have been reported.
 
The first single-dish detection of the $^{29}$SiO $v$=0 $J$=1--0 
line in IRC\,+10011 was performed by Cho et al. (\cite{cho}). 
Subsequent studies of this circumstellar maser transition
 revealed that it has some properties usually associated to 
the $^{28}$SiO maser lines, such as its time variability or the
correlation with the 8\,$\mu$m stellar radiation 
(Alcolea \& Bujarrabal 1992).

In the following sections we present the first map of the $^{29}$SiO 
$v$=0 $J$=1--0 maser line in an evolved star, IRC\,+10011, 
 using very long baseline interferometry. We compare the results
 obtained for this transition with those of the $^{28}$SiO lines,
 and we also discuss how our observational results may affect 
the $^{29}$SiO circumstellar maser theory.

\section{Observations and data analysis}

The observations of the SiO maser transitions  in
IRC\,+10011 were performed on 2002 December 7 using the
NRAO\footnote{The National Radio Astronomy Observatory is a facility
of the National Science Foundation operated under cooperative
agreement by Associated Universities, Inc.} Very Long Baseline Array
(VLBA).
We observed the $^{29}$SiO $v$=0 $J$=1--0,  $^{28}$SiO $v$=1 
$J$=2--1, $^{28}$SiO $v$=1 $J$=1--0 and the $^{28}$SiO $v$=2 
$J$=1--0 (whose rest frecuencies are 42879.916, 86243.442, 43122.080
and 42820.587 MHz respectively).

These observations correspond to the second epoch of a VLBA 
multi-epoch/transitional study of some SiO maser lines in a 
sample of AGB stars. All the transitions were detected and 
 mapped. We note that for the first time, we have been able 
to map the $^{29}$SiO emission in a circumstellar envelope. 

The data were correlated at NRAO facilities in Socorro (New
Mexico). The \mbox{43 GHz} transitions were recorded simultaneously
and separated a few hours to the 86 GHz data.  The spectral
resolutions achieved were 0.22 \,km\,s$^{-1}$ and 0.11 \,km\,s$^{-1}$
using a 8 MHz and 16 MHz bandwidths. The calibration was done using
the Astronomical Image Processing System (AIPS) package, following the
standard procedures for spectral line VLBI observations.

\begin{figure}
  \centering \includegraphics[angle=0, width=0.4\textwidth]
   {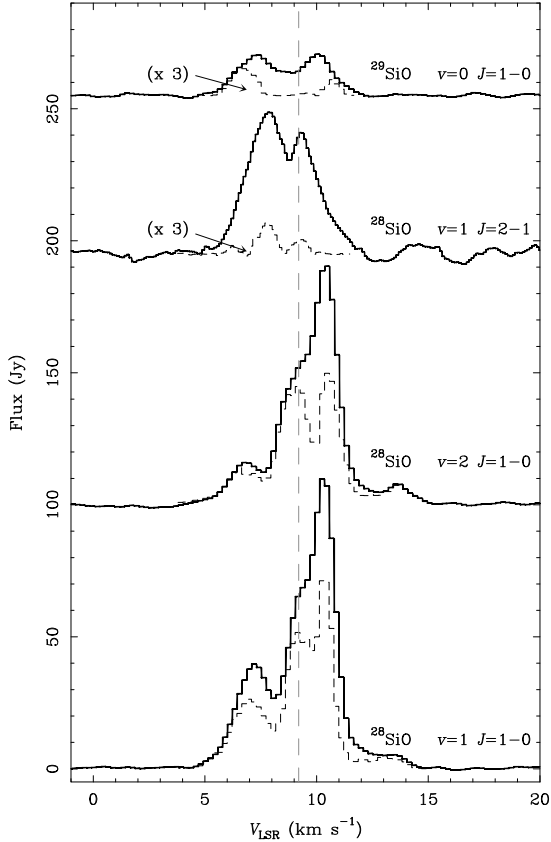}
  \caption{The total power (continuous lines) and recovered flux
 (dashed lines) spectra in IRC\,+10011. 
  The dashed grey line indicates the systemic velocity of the source.}
  \label{espec}
  \end{figure}

\section{Results}

We present in Fig.\,\ref{espec} the total power and the recovered maser
intensity, after full calibration and imaging, of the different 
observed transitions. The single-dish spectra are composed of
multiple peaks with velocities near the systemic velocity of the
source ($V_{LSR}$= 9.2 km\,s$^{-1}$, from Cernicharo et al.
\cite{cerni}). The line profiles of the $v$=1 and $v$=2 $J$=1--0
masers are similar although they differ from the $v$=1 $J$=2--1 and
$^{29}$SiO $v$=0 $J$=1--0 ones.
 
\begin{figure*}[!ht] 
 \vspace{8cm} 
\hspace{-1.5cm}
\includegraphics{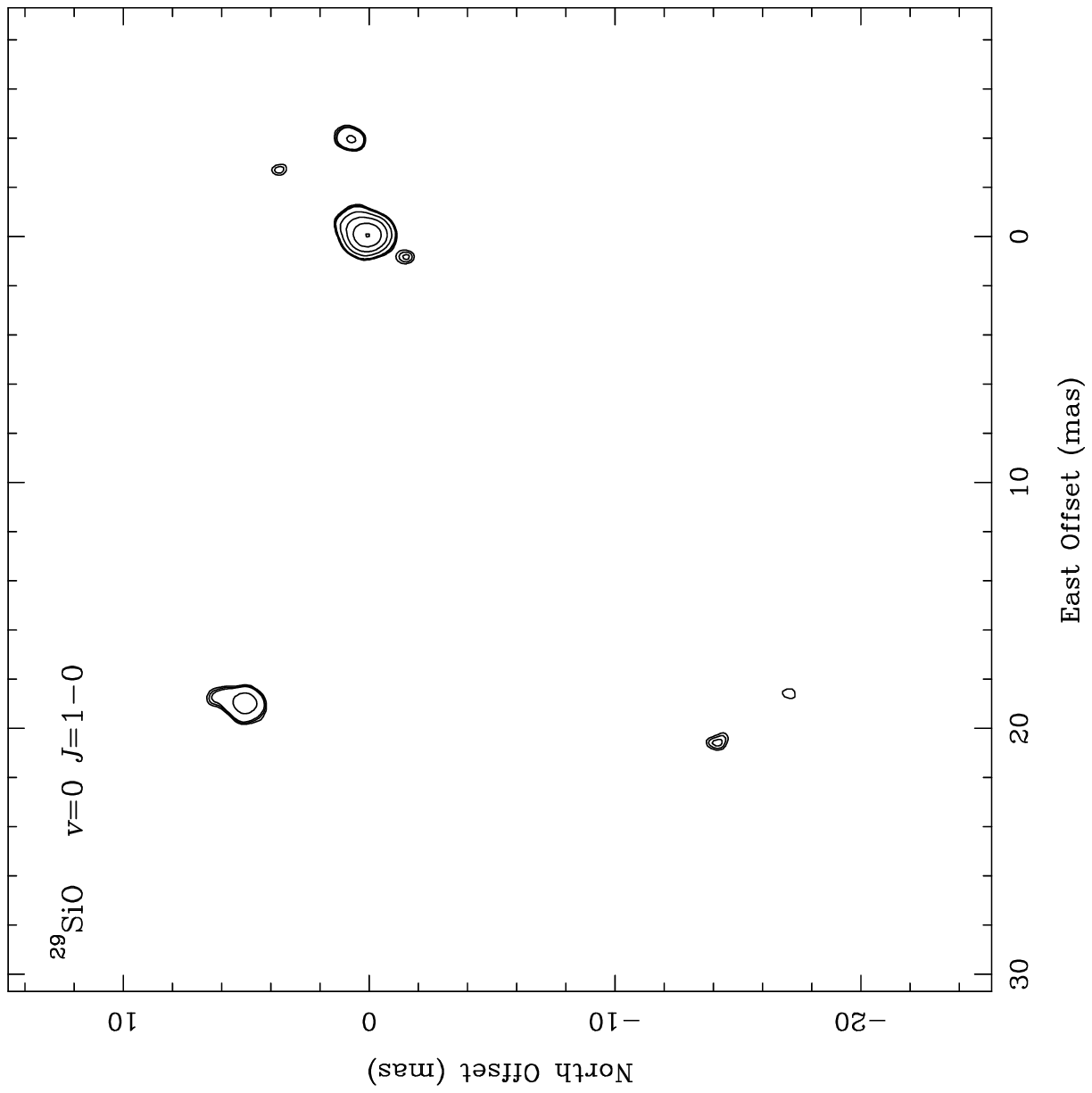} \includegraphics{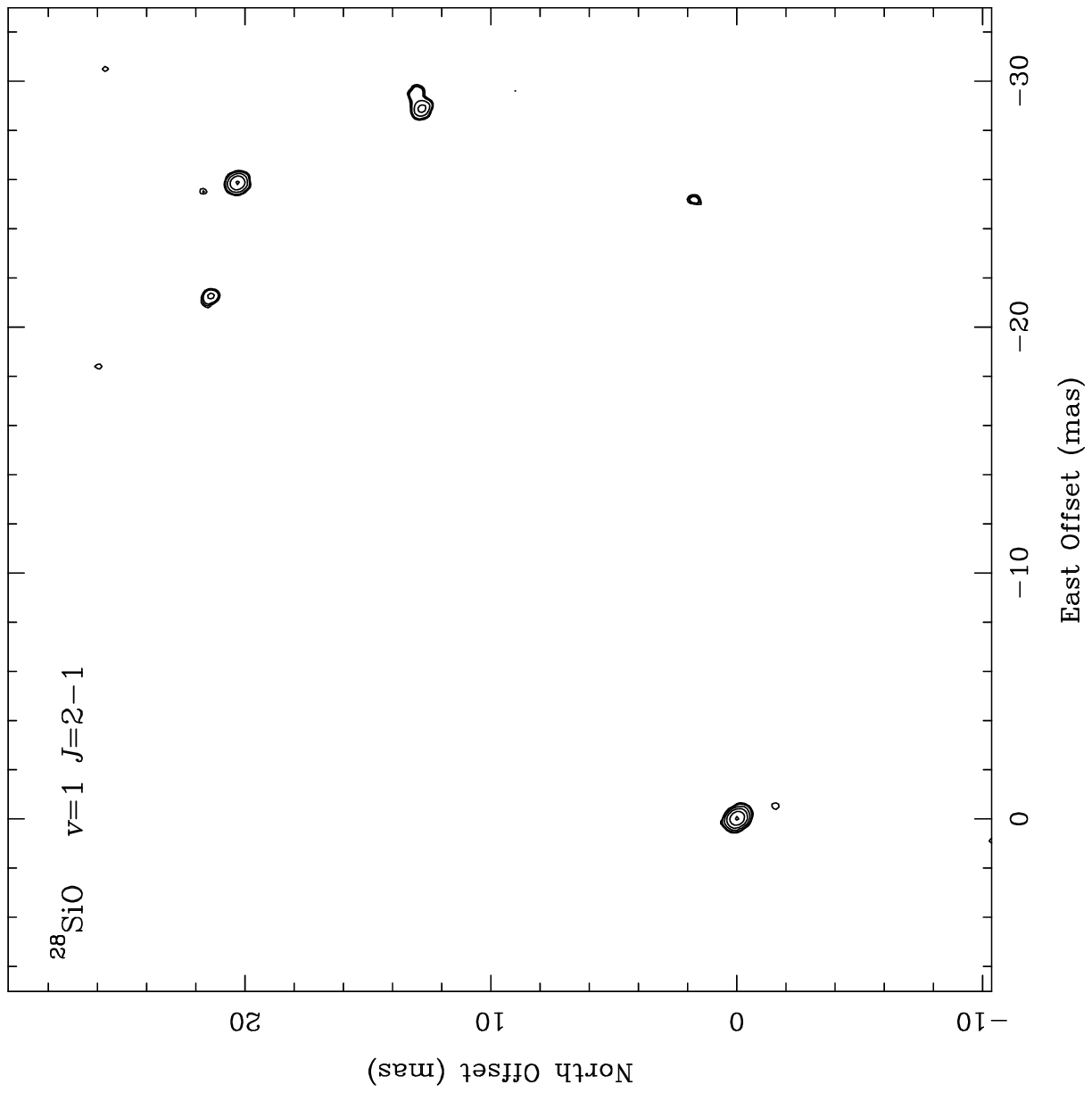}
\vspace{2.2cm}
\caption{Integrated intensity maps for the $^{29}$SiO $v$=0 $J$=1--0
 (left) and $^{28}$SiO $v$=1 $J$=2--1 (right) masers in IRC\,+10011. 
  The peak flux is 1.77 and 2.41 Jy beam$^{-1}$\,$\cdot$\,km\,s$^{-1}$
  respectively. The contour levels are 0.08, 0.10, 0.12, 0.23, 0.46, 
  0.92 and 1.76 Jy beam$^{-1}$ $\cdot$\,km\,s$^{-1}$ for the $J$=1--0 
  and 0.10, 0.12, 0.14, 0.28, 0.57, 1.14 and 2.28 Jy beam$^{-1}$ 
  $\cdot$\,km\,s$^{-1}$ for the $J$=2--1.}
\label{29y3} 
\end{figure*}

\begin{figure*}[!ht]
\vspace{8cm} 
\hspace{-1.5cm}\includegraphics{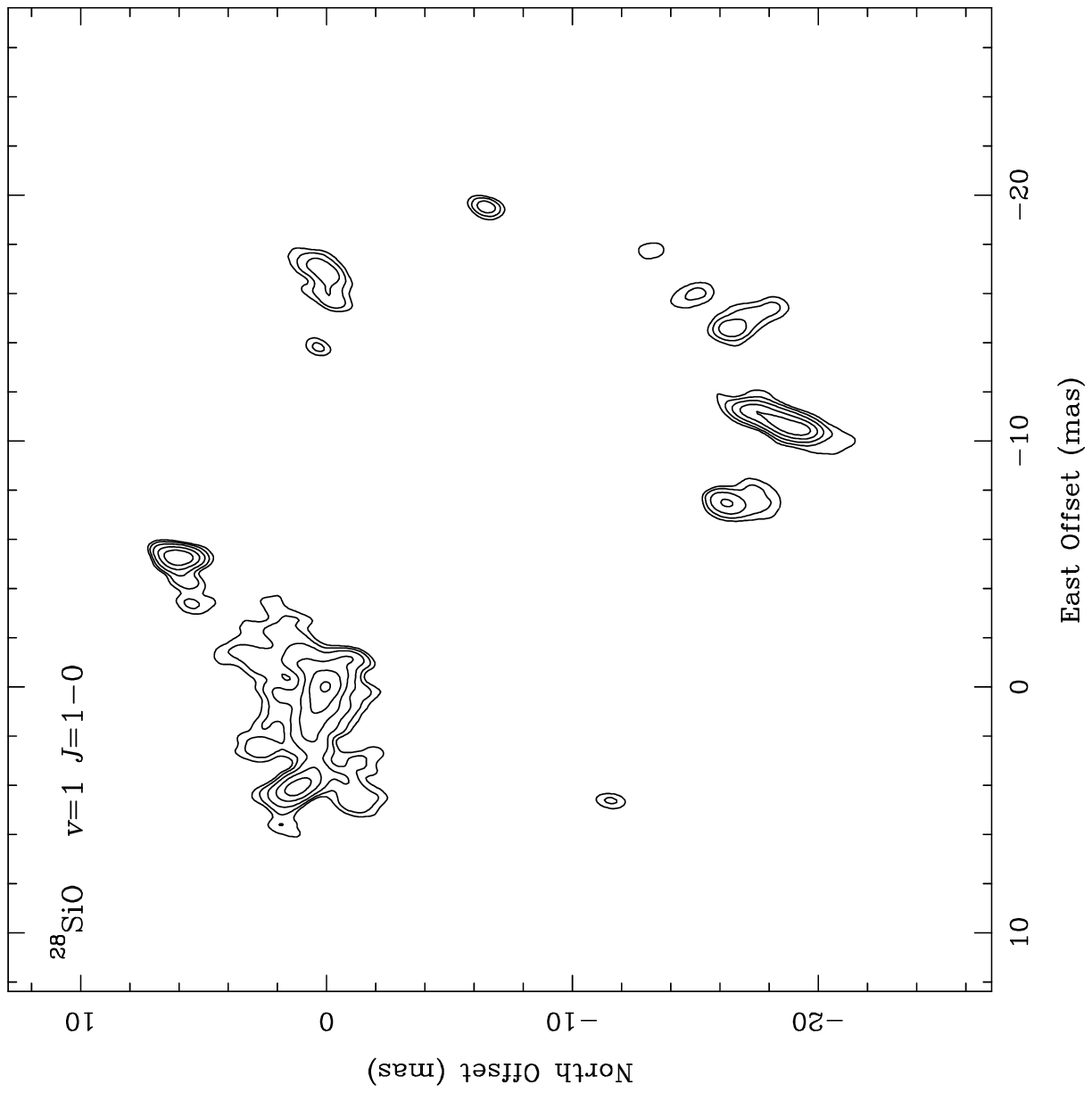} \includegraphics{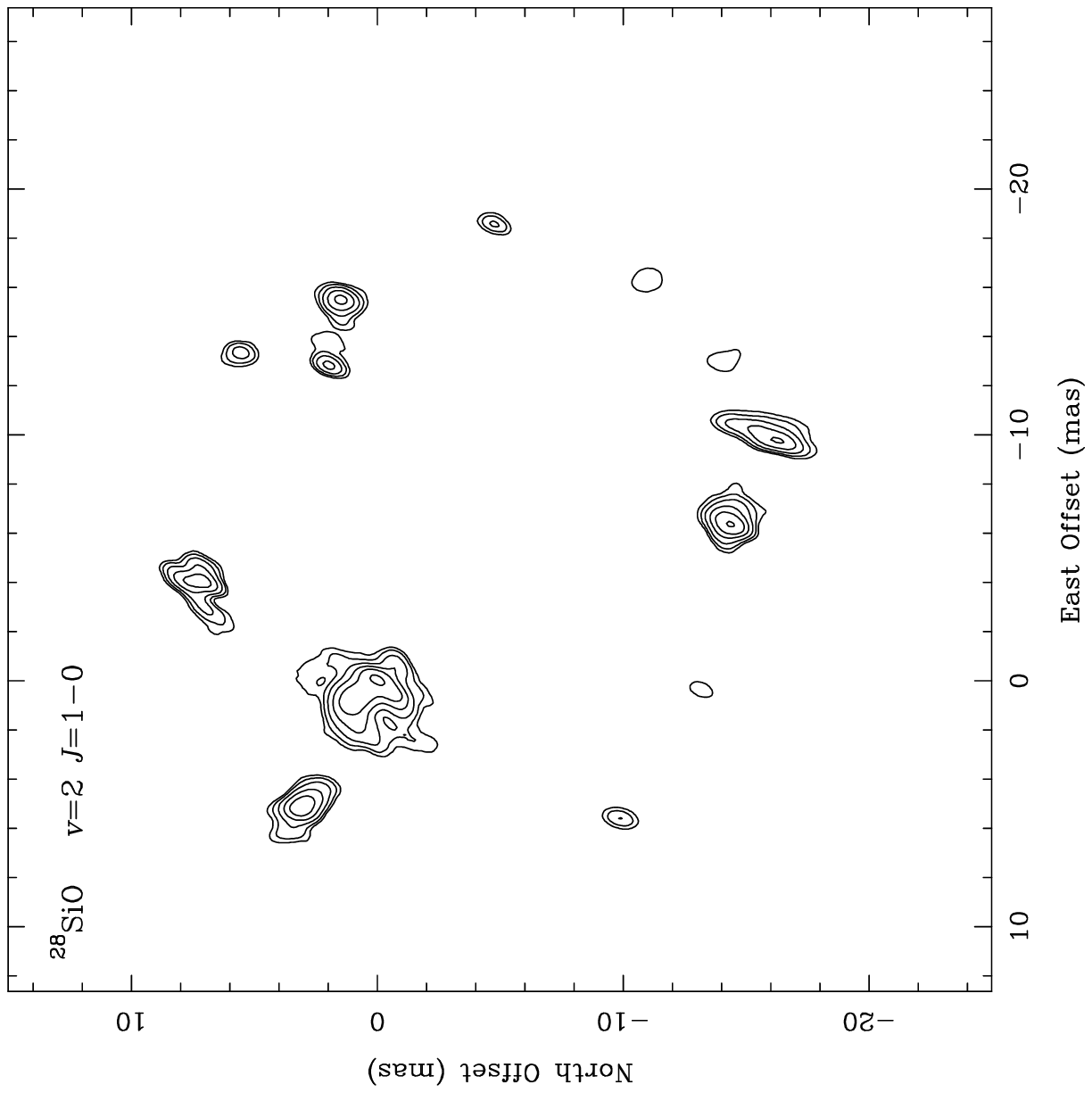}
\vspace{2.2cm}
\caption{Integrated intensity maps for the $^{28}$SiO $v$=1 (left) and $v$=2 
  (right) $J$=1--0 rotational transitions in IRC\,+10011. The peak
  flux is 9.14 and 9.72 Jy beam$^{-1}$\,$\cdot$\,km\,s$^{-1}$
  respectively. The contour levels in both images are 0.28, 0.57,
  1.14, 2.28, 4.16 and 8.34 Jy beam$^{-1}$\,$\cdot$\,km\,s$^{-1}$.  }
\label{v1v2} 
\end{figure*}

 The integrated intensity maps are shown in Figs. \ref{29y3} and
 \ref{v1v2}.  The  $^{29}$SiO $v$=0 $J$=1--0 and $^{28}$SiO $v$=1 
$J$=2--1 maser maps are composed of a few spots forming an incomplete
 ring. In contrast, the ring-like geometry is clear for 
the $v$=1 and $v$=2 $J$=1--0 lines.  
 The map of the $^{29}$SiO transition (left panel of Fig. \ref{29y3})
is composed of 7 maser spots with velocities ranging from 5.7 to 
11.3 km\,s$^{-1}$. The brightest feature has an intensity of 
$\sim$1.76 Jy beam$^{-1}$ $\cdot$\,km\,s$^{-1}$. The
$^{29}$SiO emission is the weakest among the four lines observed.
 The emission forms a ring, though incomplete, with a mean 
radius of $\sim$13.5 mas, therefore this maser radiation is probably
 amplified tangentially, as other well studied maser transitions.

\begin{figure*}[]
 \vspace{8cm} 
\hspace{-1.5cm}\includegraphics{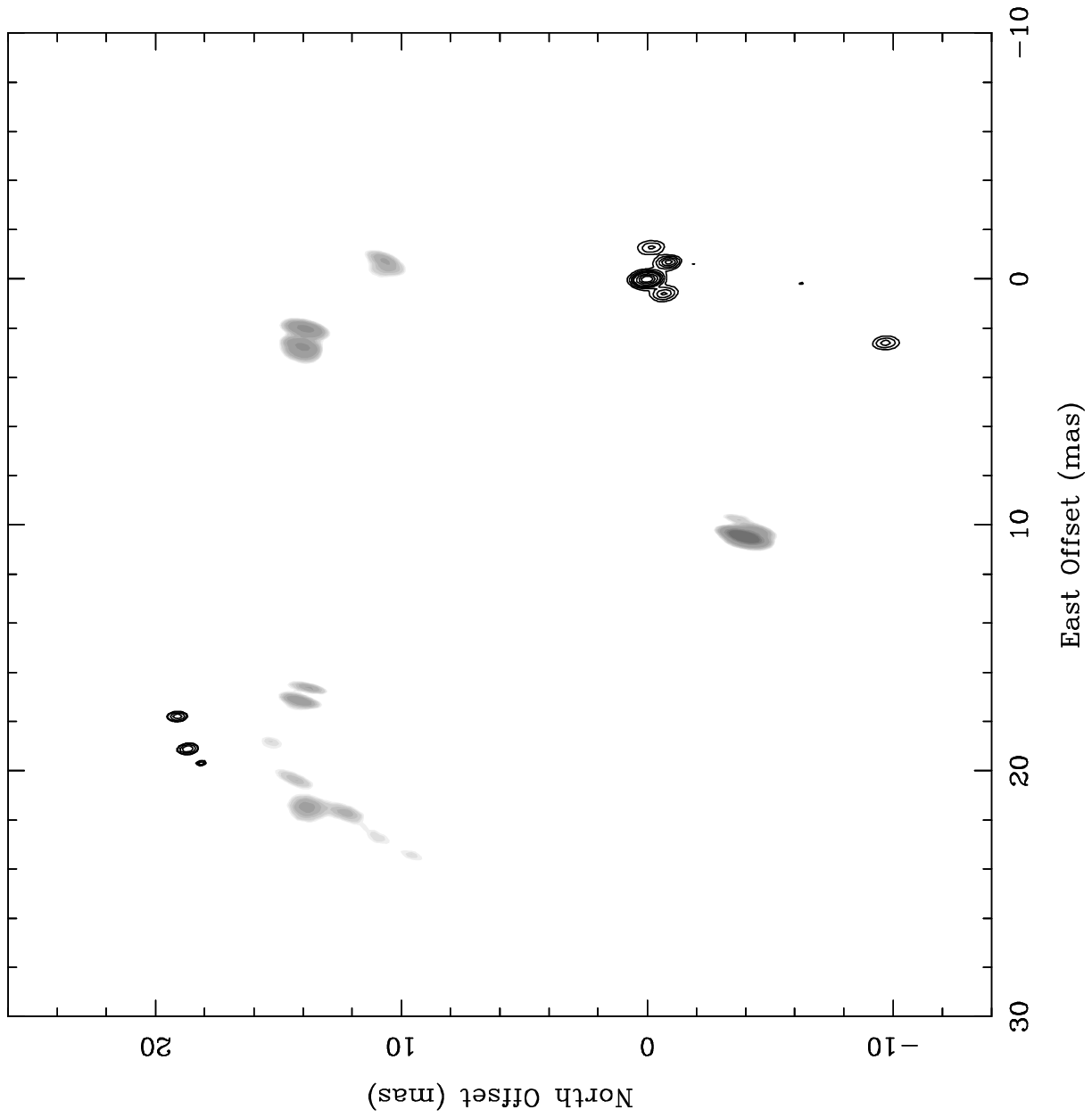} \includegraphics{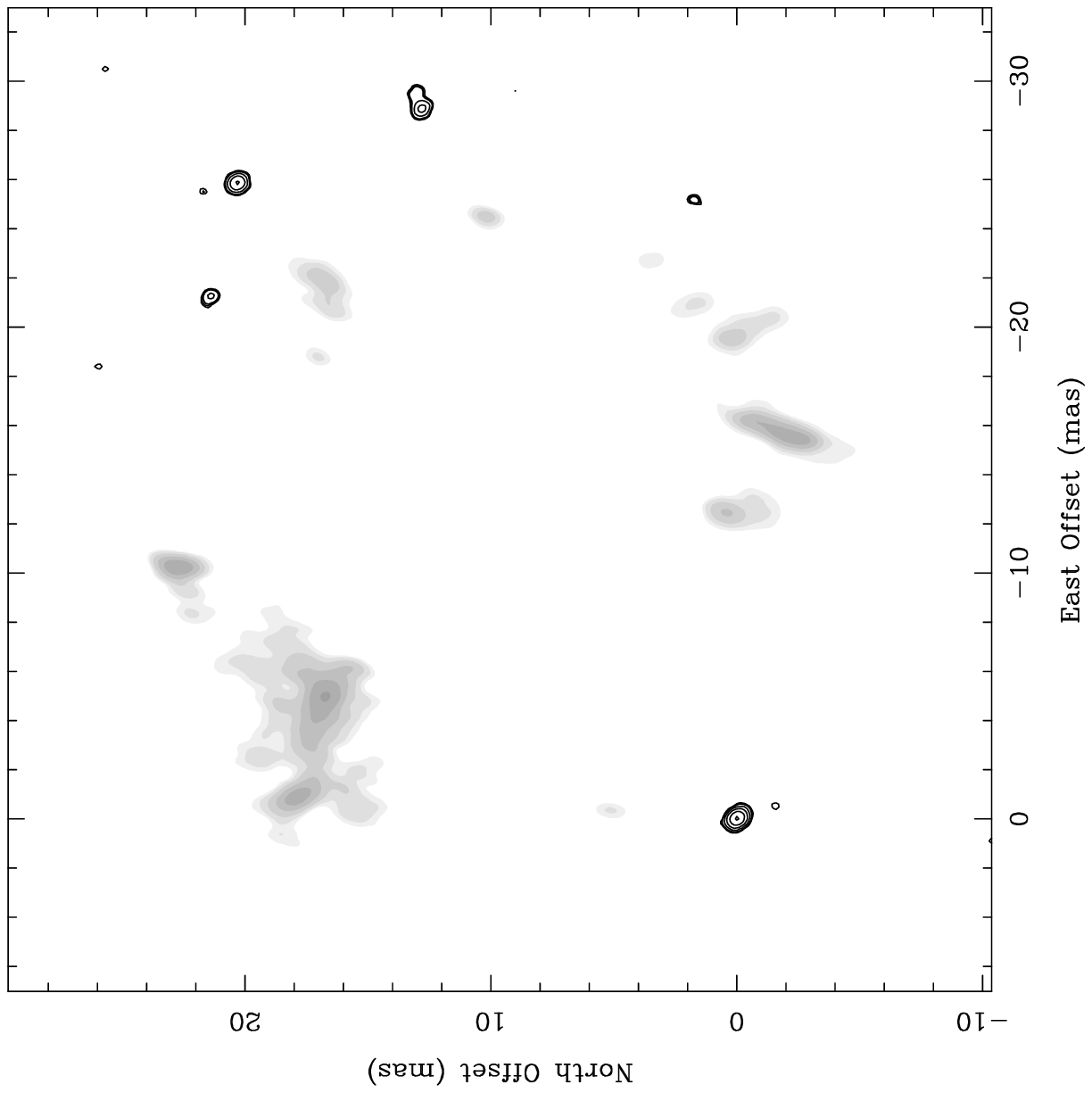}
\vspace{2.2cm}
\caption{Comparison of the $^{28}$SiO $v$=1 $J$=1--0 (grey) and 
$J$=2--1 (line contours) rotational transitions in IRC\,+10011.
  Left: results from  Soria-Ruiz et al. (\cite{soriaruiz}). 
Right: results of this work.}
 \label{comps} 
 \end{figure*}

From the comparison of rotational transitions within the same
vibrational state, the $^{28}$SiO $v$=1 $J$=1--0 and $J$=2--1, we
obtained that the latter line is produced in a layer further away
than the $J$=1--0 one. Furthermore, the $^{29}$SiO $v$=0 $J$=1--0
 masing region appears to be located in a layer in between these 
two transitions.  In this case, the spatial distributions of the
$v$=1 regions clearly differ (right and left panels of Figs. 
\ref{29y3} and \ref{v1v2}). 
This result is compatible with that obtained in previous 
observations of IRC\,+10011 (see \mbox{Fig.\ref{comps})}. 

The $v$=1 and $v$=2 $J$=1--0 $^{28}$SiO lines (Fig. \ref{v1v2}) 
present ring structures composed of multiple spots with similar
distributions. In IRC\,+10011, Desmurs et al. (\cite{desmurs}) and
Soria-Ruiz et al. (2004) found a systematic shift between these two
lines of about 1--3 mas, being the $v$=2 always located in an inner
region of the envelope. We confirm this result since our $v$=1 and
$v$=2 maps are displaced $\sim$1 mas (see Fig. \ref{v1v2}).  We note
that similar angular shifts have also been measured in other Mira
variable stars (Cotton et al. \cite{cotton}).

Furthermore, although the observational results are consistent with
those obtained in the first epoch, there are some small differences 
when comparing the maps. For example, the \mbox{86 GHz} $^{28}$SiO $v$=1 
$J$=2--1 emission is distributed in different regions of the 
 circumstellar envelope (see both panels of Fig. \ref{comps}).
 However, this in not the case of the 43 GHz masers, the 
$^{28}$SiO $v$=1 $J$=1--0 and $^{28}$SiO $v$=2 $J$=1--0, which show in
 both epochs some maser features in common.

\section{Discussion}

Some theoretical models have been proposed to explain the $^{29}$SiO
$v$=0 $J$=1--0 maser amplification in these evolved stars.  Robinson
\& Van Blerkom (\cite{robin}) and Deguchi \& Nguyen-Quang-Rieu
(\cite{deguchi}) suggested that this ground-state masers were produced
if the vibrational transitions present a significantly higher opacity
along the radial direction than in the tangential one, predicting a
two-peaked line profiles similar to the circumstellar OH lines.
However, the model does not reproduce the observations since the line
shapes of $^{29}$SiO masers are composed of narrow peaks near the
stellar velocity (Alcolea \& Bujarrabal \cite{alcolea}), 
and, in addition, our maps indicate a ring-like geometry, very
probably due to tangential amplification.

Other physical mechanisms involved in the SiO excitation are 
the line overlaps between infrared transitions of $^{28}$SiO and
$^{29}$SiO. Those were firstly suggested by Olofsson et al. 
(\cite{olofsson}) to explain the $v$=0 $^{29}$SiO maser lines. 
Subsequent calculations by Cernicharo et al.  (\cite{cerni}) 
indicate that the $v$=0 $J$=1--0 $^{29}$SiO line might be 
effectively pumped by the overlap between the $^{28}$SiO $v$=2 
$J$=4 $\rightarrow$ $v$=1 $J$=3 and the $^{29}$SiO $v$=1 $J$=0 
$\rightarrow$ $v$=0 $J$=0 ro-vibrational transitions.

The fact that the $v$=0 $J$=1--0 $^{29}$SiO maser is confined in a
region of the envelope in between the $^{28}$SiO $v$=1 $J$=1--0 
 and \mbox{$J$=2--1} masing shells, indicates that this maser line 
requires high excitation temperatures (T$_{ex}$\,$\sim$\,1700 K).
 This can be explained if the inversion population of the ground
 state is mainly produced via de-excitations from the $v$\,$>$\,0
 vibrational levels.
In addition, one possible process to pump the $^{29}$SiO $v$=0 
$J$=1--0 line could be the frequency overlap with different ro-vibrational
 transitions of abundant molecules, such as $^{28}$SiO, H$_{2}$O,
 or others.

However, none of these theories explain how their excitation
mechanisms affect the spatial distribution of the different 
$^{28}$SiO and $^{29}$SiO maser lines. Therefore, this 
makes very difficult to compare our observational results with the 
predictions of the proposed $^{29}$SiO pumping models.

This type of observations,
where multiple SiO maser transitions are measured, allow us to compare 
directly regions of the envelope with different physical conditions. 
 Moreover, from these presented VLBA images, we can see that the 
width of the whole masing shell is considerable ($\sim$\,7-9 mas) and 
may not be as thin as that obtained from the observations of 
a single maser line.  
 
\section{Summary and work in progress}

We have presented the first image of the $^{29}$SiO $v$=0 $J$=1--0
maser line in a late type star, IRC\,+10011. The distribution of the
emission suggests that, as in other well studied circumstellar masers,
the amplification is tangential.  The maser shell is ring-like, and
located in a region in between the $^{28}$SiO $v$=1 $J$=2--1 and $v$=1
$J$=1--0 rings.

We confirm that the $v$=1 $J$=2--1 maser (86 GHz) in IRC\,+10011 
is formed further away from the central AGB star, and that the $v$=1
 and $v$=2 $J$=1--0 (43 GHz) masing regions have a similar 
distribution, though clearly not identical.

At present we are analizing the same observed SiO masers in other
variable stars: R\,Leo, TX\,Cam and $\chi$\,Cyg. We are also
introducing the line overlaps in the theoretical pumping calculations
of the $^{29}$SiO emission to study the possible physical processes
involved in this peculiar amplification effect.

\begin{acknowledgements}
  This work has been financially supported by the Spanish DGI (MCYT)
  under projects AYA2000-0927 and AYA2003-7584, and by the European 
  Commission's I3 Programme ``RADIONET", under contract No.\ 505818.
\end{acknowledgements}

\end{document}